# Efficiently charting the space of mixed vacancy-ordered perovskites by machine-learning encoded atomic-site information

Fan Zhang[1], Li Fu[1], Weiwei Gao[1*], Peihong Zhang[2], Jijun Zhao[3*]

1. *Key Laboratory of Material Modification by Laser, Ion and Electron Beams (Dalian University of Technology), Ministry of Education, Dalian 116024, China*
2. *Department of Physics, State University of New York at Buffalo, Buffalo, New York 14260, USA*
3. *Guangdong Basic Research Center of Excellence for Structure and Fundamental Interactions of Matter, Guangdong Provincial Key Laboratory of Quantum Engineering and Quantum Materials, School of Physics, South China Normal University, Guangzhou 510006, China*

**Abstract**

Vacancy-ordered double perovskites (VODPs) are promising alternatives to three-dimensional lead halide perovskites for optoelectronic and photovoltaic applications. Mixing these materials creates a vast compositional space, allowing for highly tunable electronic and optical properties. However, the extensive chemical landscape poses significant challenges in efficiently screening candidates with target properties. In this study, we illustrate the diversity of electronic and optical characteristics as well as the nonlinear mixing effects on electronic structures within mixed VODPs. For mixed systems with limited local environment options, the information regarding atomic-site occupation in-principle determines both structural configurations and all essential properties. Building upon this concept, we have developed a model that integrates a data-augmentation scheme with a transformer-inspired graph neural network (GNN), which encodes atomic-site information from mixed systems. This approach enables us to accurately predict band gaps and formation energies for test samples, achieving Root Mean Square Errors (RMSE) of 21 meV and 3.9 meV/atom, respectively. Trained with datasets that include (up to) ternary mixed systems and supercells with less than 72 atoms, our model can be generalized to medium- and high-entropy mixed VODPs (with 4 to 6 principal mixing elements) and large supercells containing more than 200 atoms. Furthermore, our model successfully reproduces experimentally observed bandgap bowing in Sn-based mixed VODPs and reveals an unconventional mixing effect that can result in smaller band gaps compared to those found in pristine systems.

**Introduction**

Fabricating mixtures of electronically distinct yet structurally compatible materials represents a significant paradigm for tailoring material properties, garnering considerable attention in both fundamental material research and practical applications. The perovskite structure is renowned for its structural and chemical flexibility, capable of accommodating a diverse array of elemental combinations, thereby creating an extensive chemical design space[1, 2, 3, 4]. Vacancy ordered double perovskites (VODPs), with a general chemical formula of $A_2MX_6$ (A = K, Rb, Cs, NH$_4$; M = Ti, Zr, Pd, Sn, Te, Hf, Pt; X = Cl, Br, I) are quasi-zero-dimensional perovskites. Free of the toxic lead element, VODPs are environmentally friendly alternatives to conventional halide perovskites $AMX_3$ (A = FA, MA, Cs; M = Pb, Sn; X = Br, Cl, I) for their potential applications in photovoltaic and light-emitting devices and have been widely studied[1, 2, 3, 4, 5, 6]. Experiments have also demonstrated the excellent miscibility of VODPs[3, 7, 8, 9, 10]. Notably, Folgueras. et al. have recently demonstrated room-temperature-solution and low-temperature-solution procedures for synthesizing a class of high-entropy mixed VODPs with up to six principal components at the M-site[11].

Explorations of the chemical space of mixed systems have been mostly led by low-throughput experiments and sometimes followed by theoretical and computational investigations. Development of more accurate and efficient computational methods for high-throughput computational screening of mixed material systems can greatly expedite materials design and discovery process. The challenge mainly stems from the heavy cost of first-principles computation to sample the nearly infinite possibilities of random configurations in mixed structures. There are two widely exploited strategies, effective-medium methods, such as the virtual crystal approximation (VCA)[12] and the coherent potential approximation (CPA)[13] methods, as well as supercell-based methods, such as the special quasi-random structure (SQS)[14] approach, the similar local atomic environment (SLAE)[15], the cluster expansion (CE)[16, 17] and similar atomic environment (SAE)[18] method. However, the effective-medium methods tend to ignore the effects of local environments[19, 20, 21, 22], while the supercell-based methods typically require time-consuming first-principles modeling of large supercells with hundreds of atoms[23, 24, 25, 26, 27, 28] to account for the random distribution of atoms in multicomponent mixed systems. All these difficulties call for the development of computational methods that are significantly more efficient than traditional first-principles approaches for addressing large and complex multi-component supercells.

In the past decade, machine learning (ML) has become an influential tool for expediting high-throughput screening of mixed systems such as mixed halide perovskites[29]. For example, Choubisa et al. proposed a ML method based on "crystal site feature embedding" representation for property prediction of mixed halide perovskites $MAPb(I_xBr_{(1-x)})_3$ and $MAPb_xSn_{(1-x)}I_3$[30]. This model achieves a Mean Absolute Error (MAE) of 69 meV for band gap prediction. Arun et al. developed a data-driven framework for property prediction of halide perovskite alloys with up to four B-site elements. They achieved a Root Mean Square Errors (RMSE) of 220 meV for band gap prediction[31]. Recently, Kim et al. employed a crystal graph convolutional neural network (CGCNN) to investigate the decomposition energies and bandgaps of 41,400 mixed $ABX_3$ metal halide perovskites with up to four B-site elements. The CGCNN model has an MAE of 37 meV for the bandgap prediction[32]. Despite the significant advancements brought by the above methods, it is still a major challenge to generalize ML model to predict the physical property of medium- or high-entropy systems of more than four elements and multi-site mixing, which are key for exploring high-entropy mixed semiconductors or alloys.

Using mixed VODP systems as examples, herein we have developed an ML model that hold great promise for effective handling of medium- and high-entropy mixed perovskite materials. We first demonstrate the exceptional tunability of the electronic and optical properties of mixed VODP using density functional theory (DFT) and many-body perturbation theory (MBPT). These initial results then serve as training datasets for a transformer-inspired GNN model for efficient prediction of the properties of VODPs. Our GNN model only utilizes atomic-site occupation information of a mixing structure as input, and can be generalized for predicting large and high-entropy systems with more than 200 atoms and with 4 ~ 6 mixing elements, respectively.

# Results and discussion

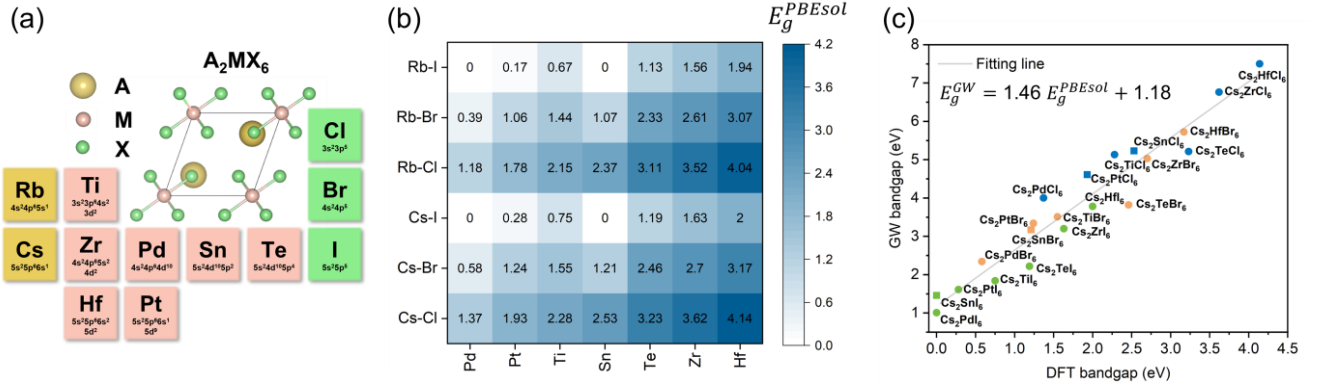

Fig. 1. (a) The crystal structure of $A_2MX_6$ within a primitive cell and the elements list for A, M and X-sites considered in this work. (b) Color map of DFT band gaps for pure $A_2MX_6$, with A = Rb, Cs; M = Pd, Pt, Ti, Sn, Te, Zr, Hf; and X = Cl, Br, I. (c) The relation between GW and DFT band gaps for $Cs_2MX_6$, with M = Pd, Pt, Ti, Sn, Te, Zr, Hf; and X = Cl, Br, I. Rectangles and spheres indicate direct and indirect band gaps, respectively. The solid line shows the fitting function of the data points.

**Diverse and highly tunable electronic structures and optical properties**

To begin with, we consider pure VODPs $A_2MX_6$ with A = Rb, Cs; M = Ti, Zr, Pd, Sn, Te, Hf, Pt; and X = Cl, Br, I, which have been synthesized experimentally[1, 3, 33, 34, 35, 36, 37, 38, 39]. As highlighted in Fig. 1b, the DFT band gaps of pure VODPs calculated with PBEsol exchange-correlation functional span from 0 to 4.14 eV. To accurately evaluate the quasiparticle band gaps, we employed many-body perturbation theory calculations based on the one-shot GW approximation, i.e., $G_0W_0$. As plotted in Fig. 1c, the GW quasiparticle band gaps of $A_2MX_6$ vary from 1.0 to 7.5 eV and exhibit roughly linear correlation with the DFT band gaps. Generally, we note X-site and M-site elements prominently affect the frontier electronic states, while the impact of A-site elements to the band gap is relatively small (less than 0.2 eV). The band structures of $Cs_2MX_6$ are depicted in Fig. S1-S3. Our DFT calculations suggest that the nature of band gap is highly correlated with the M-site elements. For instance, Sn-based VODPs generally have direct band gaps; VODPs with M-site = Te, Ti, Zr, Pd, and Hf have indirect band gaps. Notably, the minimum direct band gaps of Ti-, Zr- and Hf-based VODPs are just slightly larger (by 0.1 eV) than the indirect band gaps. In addition to the band gaps, the effective masses of these materials also demonstrate high tunability, ranging from 0.17 to $3.31 m_e$ ($m_e$ is electron mass) as summarized in Table S1.

The frontier electronic states derived from the X- and M-sites significantly influence the optical properties of VODPs, showcasing a rich array of excitonic physics within these materials. We illustrate this point by analyzing the lowest-energy exciton wavefunctions of $Cs_2TiI_6$, $Cs_2TeI_6$ and $Cs_2SnI_6$, as depicted in Fig. 2. These prototypical VODPs have distinct effective masses of electron and hole with a quasi-direct band gap, indirect band gap, and direct band gap for $Cs_2TiI_6$, $Cs_2TeI_6$ and $Cs_2SnI_6$, respectively. Our findings indicate that the lowest-energy exciton is dark in $Cs_2TiI_6$ and $Cs_2SnI_6$ but bright in $Cs_2TeI_6$. As shown in Fig. 2a to 2c, the wavefunction of the lowest-energy exciton becomes increasingly delocalized going from $Cs_2TiI_6$ to $Cs_2TeI_6$ and then to $Cs_2SnI_6$, suggesting a transition from Frenkel-like to Wannier-Mott-like excitons. By fixing the hole position at an iodine atom, we observe that electron distribution is primarily confined within a single $[TiI_6]^{2-}$ octahedron in $Cs_2TiI_6$, but becomes more extended in both $Cs_2TeI_6$ and $Cs_2SnI_6$. Furthermore, we analyze the Brillouin Zone (BZ) distribution of the lowest-energy exciton by considering their electron-hole amplitude $A_{cv\mathbf{k}}$, where $v$, $c$, and $\mathbf{k}$ are the indices of valence, conduction, and $k$-points. As illustrated in Fig. 2d-f, the sizes of colored circles are proportional to $\sum_v |A_{cv\mathbf{k}}|^2$ for conduction states $c$ and $\sum_c |A_{cv\mathbf{k}}|^2$ for valence bands $v$ located at point $\mathbf{k}$. The results demonstrate that for $Cs_2TiI_6$, the exciton exhibits greater extension across the Brillouin zone; it becomes concentrated near high symmetry points X and L for $Cs_2TeI_6$; while it localizes predominantly around Γ for $Cs_2SnI_6$.

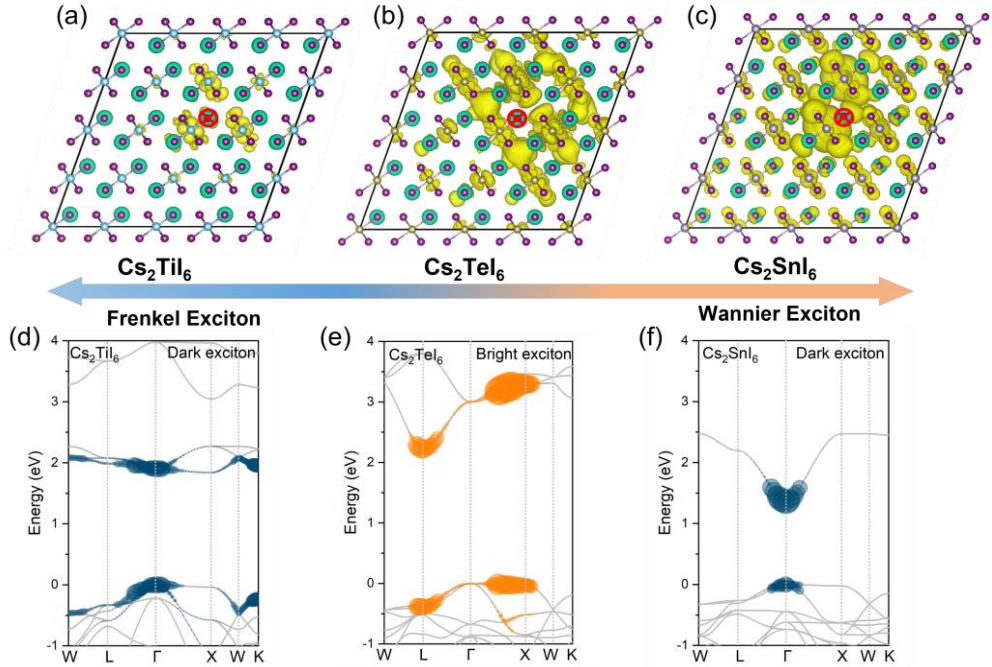

Fig. 2. Real-space distribution of electrons for (a) the lowest-energy dark exciton of $Cs_2TiI_6$, (b) the lowest-energy bright exciton of $Cs_2TeI_6$ and (c) the lowest-energy dark exciton of $Cs_2SnI_6$. The hole position is fixed at an iodine atom, which is marked by a red cross. About 80% of the electron density is within the isosurface (shown in yellow). Reciprocal-space band distribution of exciton wavefunctions of (d) $Cs_2TiI_6$, (e) $Cs_2TeI_6$ and (f) $Cs_2SnI_6$.

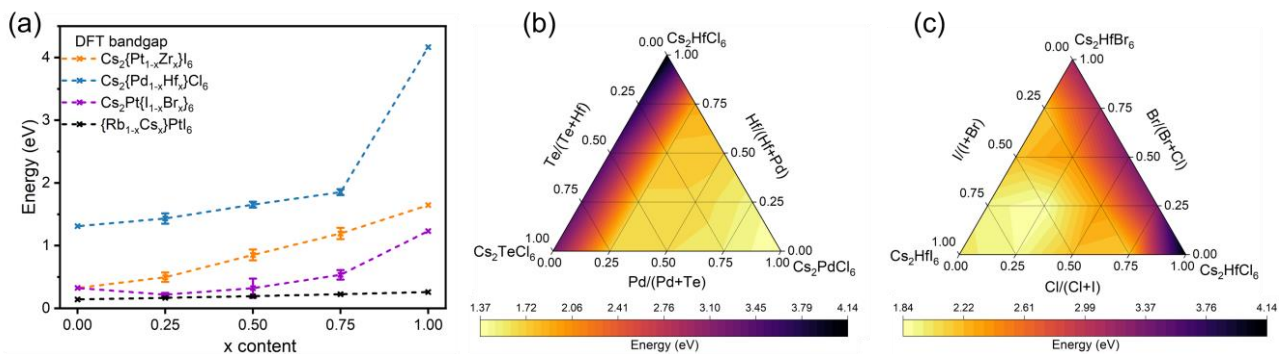

Fig. 3. (a) The variation of DFT band gap with different mixing ratios. (b) DFT band gap of $Cs_2\{HfTePd\}Cl_6$ and (c) $Cs_2Hf\{ClBrI\}_6$.

**Nonlinear mixing effects in mixed VODPs**

The diversity of A-site elements and $[MX_6]^{2-}$ octahedra in VODP structures facilitates precise tuning of physical properties through compositional mixing. In this study, we concentrate on the effects of mixing on electronic band gap and formation energy. Figure 3a illustrates the variation in band gap resulting from binary mixing at the A, M, and/or X-sites. Generally speaking, compared with the mixing of A-site elements, alterations in band gap are more pronounced when M- or X-site elements are mixed. The band gaps of $Cs_2\{Pt_{1-x}Zr_x\}Cl_6$ and $\{Rb_{1-x}Cs_x\}PtI_6$ change nearly linearly with the mixing ratio $x$ (Fig. 3a). On the other hand, the band gaps of some M- or X-site mixed VODP materials, such as $Cs_2\{Pd_{1-x}Hf_x\}Cl_6$ and $Cs_2Pt\{I_{1-x}Br_x\}_6$ show significant deviation from the linear behavior. Such nonlinear composition-dependence of band gap is known as the "bandgap bowing effect", which has been observed in Sn-based mixed halide VODPs[40] and some other mixed systems[41, 42, 43]. Additional calculations on ternary mixing systems also demonstrate more complex nonlinear mixing effects, as shown in Fig. 3b and 3c. Interestingly, the X-site mixed VODPs, such as $Cs_2Pt\{I_{1-x}Br_x\}$, can have a minimum band gap at a mixing ratio $x \approx 0.25$, as shown in Fig. 3a. Similarly, $Cs_2Hf\{Cl_{0.25}Br_{0.25}I_{0.5}\}_6$ has the minimum band gap within the $Cs_2Hf\{ClBrI\}_6$ mixed systems (see Fig. 3c). In addition to electronic band gaps, we have also investigated the element mixing effects on the formation energy and the bulk modulus, as shown in Fig. S4. Overall, the formation energy exhibits a linear relation with respect to the mixing ratio, while the bulk modulus shows highly nonlinear mixing effects.

**The Graph Neural Network model**

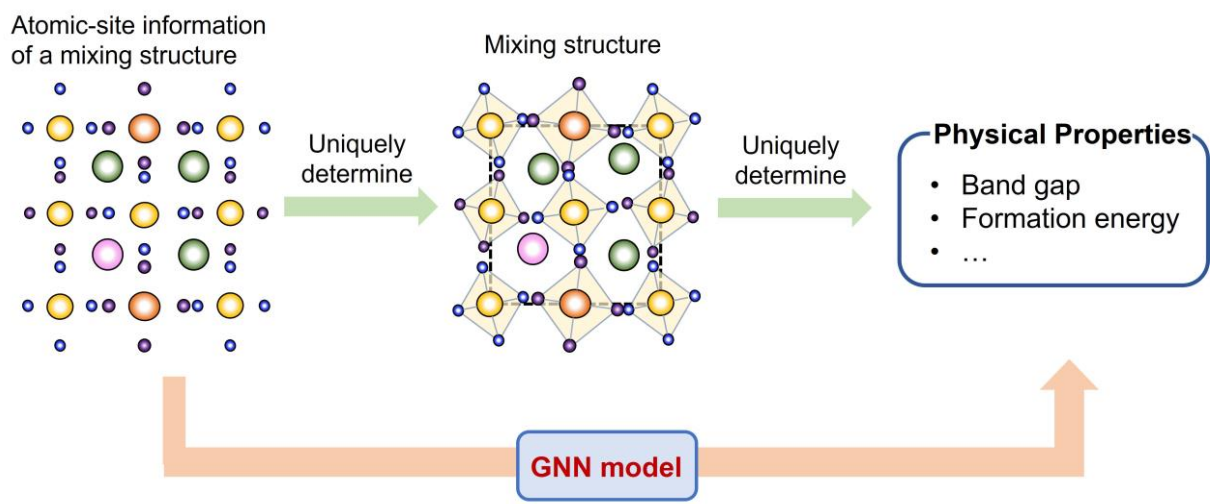

Fig. 4. Schematic for the mapping of GNN model to physical properties.

The nonlinear mixing effects complicate the design of materials and the prediction of mixed VODPs, particularly in complex systems that involve mixtures of elements at multiple sites and/or more than two principal elements for a given site. To predict the properties of such intricate mixed systems using first-principles calculations, it is typically necessary to construct a series of supercells to account for the randomness inherent in mixing configurations. A crucial observation is that, for mixed material systems with well-defined structures, both the structure and all associated properties are uniquely determined by detailed information pertaining to each atomic site. By leveraging this insight, we have developed a graph neural network (GNN) model that utilizes atomic site information from a mixed supercell as input and directly outputs physical properties. This scheme enables us to bypass both the computationally expensive structural relaxation and property evaluation steps, as illustrated schematically in Fig. 4. The following sections briefly outline the key components and motivations behind our ML models.

We construct a multi-edge graph that encodes the relative positions among mixed atomic sites.

Consider a supercell containing N atoms. In the corresponding multi-edge graph, node $i$ represents an atom along with all its periodic images. Given only the atomic-site information from an unrelaxed structure, one cannot easily identify neighboring atoms around a specific site using a threshold radius based on exact atomic distances—this requirement is typical for conventional crystal graph neural networks. Instead, within our graph model framework, an edge is established from node $j$ to node $i$ if a periodic image atom represented by node $j$ qualifies as one of the $K_i$ nearest neighbors of atom $i$. Details regarding types and numbers of neighbors considered for different sites $i$ are provided in Table 1. A schematic representation illustrating considered neighbors at A-, M-, and X-sites can be found in Fig. 5. Overall, our methodology for constructing the multi-edge graph circumvents reliance on precise bond lengths or accurate bond angles derived from optimized supercell structures.

TABLE 1. The number of nearest neighbors considered for constructing multi-edge crystal graphs of VODPs $A_2MX_6$.

| Site $i$ | $K_i$ (The number of nearest neighbors of site $i$ considered to build the crystal crystal) | Details |
|---|---|---|
| A | 12 | 12 X-site atoms |
| M | 6 | 6 X-site atoms |
| X | 5 | 4 A-site and 1 M-site atoms |

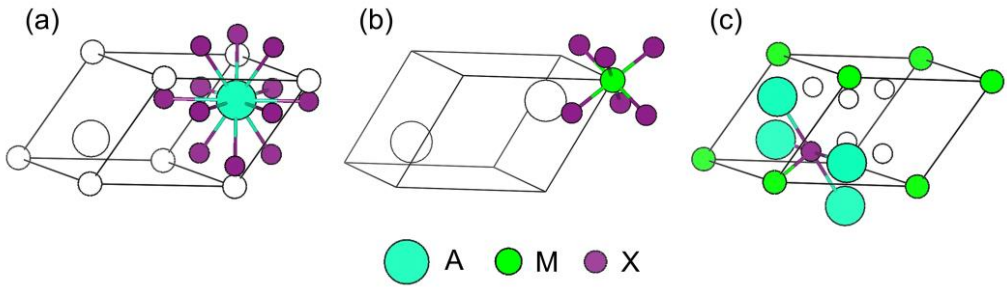

Fig. 5. Schematic for the nearest neighbors of (a) A site, (b) M site and (c) X site, which are considered for constructing the multi-edge crystal graphs of VODPs $A_2MX_6$.

To extract the physical properties from the graph representation, we implement a transformer-inspired graph layer (TGL) that integrates the invariant node-wise transformer layer[44] and the equivariant graph convolutional layer[45]. Following each TGL, we conduct standard graph mean pooling operations to aggregate the node embeddings into a cohesive graph-level descriptor vector. The graph-level descriptor is subsequently processed through a multi-layer perceptron to predict the final outcomes, such as band gap or formation energy. Further details regarding the embedding vectors and the architecture of our graph neural networks are provided in the Method section.

**Performance and applications of the GNN model**

For training the GNN model, we initially calculated DFT band gaps and formation energies of 3889 samples, which include structures with up to ternary mixed VODP systems. The dataset encompasses rhombohedral cells containing either 9 or 72 atoms and cubic supercells comprising 36 atoms. Additional information about this dataset is summarized in Table 2. Given that many unrelaxed structures converge to an identical optimized structure, our GNN model should learn a "many-to-one" mapping from unrelaxed configurations to their corresponding physical properties of the optimized structures. Therefore, to enhance our model's generalizability, we performed data augmentation by generating supplementary data samples through small random shifts (approximately 0.1 angstrom per atom) applied to the atomic positions within already generated DFT computational supercells[46]. Ninety percent of these augmented samples were designated as the training set, while the remaining ten percent were allocated as part of the test set. To further evaluate our GNN model's generalizability across larger and more complex mixed systems, we computed additional test samples that include mixed VODP supercells containing either 144 or 243 atoms along with structures featuring between four to six elements on M-site.

TABLE 2. Details of the different supercell structures of pure and mixed VODPs. $N_s$ and $N_a$ represent the number of samples and the number of atoms per cell, respectively.

| Samples | $N_s$ | $N_a$ | Details |
|---|---|---|---|
| Pure VODPs | 42 | 9 | No mixing |
| A-site mixing | 114 | 36 | Two-element mixing at A-site |
| M-site mixing | 312 | 72 | Two-element mixing at M-site |
| X-site mixing | 334 | 72/36 | Two-element mixing at X-site. |

| | | | 124 samples include 72 atoms. 210 structures include 36 atoms. |
|---|---|---|---|
| AM-site mixing | 1071 | 36 | 936 samples include two-element mixing at A- and M-sites. 135 of them include three-element mixing at A- and M-site. |
| AX-site mixing | 171 | 36 | Two-element mixing at A- and X-sites. |
| MX-site mixing | 738 | 36 | Two-element mixing at M- and X-sites. |
| AMX-site mixing | 1107 | 36 | Two-element mixing at A-, M-, and X-sites. |

The performance of our GNN model is summarized in Fig. 6. Notably, the data augmentation significantly reduces the RMSE loss, which converges as the number of augmented training samples approaches around 21,000 (see Fig. 6a). Fig. 6b and Fig. 6c compare ML-predicted and DFT-calculated results on the test set. For the prediction of band gaps, the GNN model yields an RMSE loss of 21 meV, which is well below the accuracy of typical DFT calculations. The RMSE for predicting formation energies is 3.9 meV/atom for the test dataset. Our model demonstrates superior performance compared to some recent ML models, achieving lower RMSE than the previously reported values of 146 meV for band gap prediction and 11 meV/atom for formation energy prediction in lead-free perovskites[47].

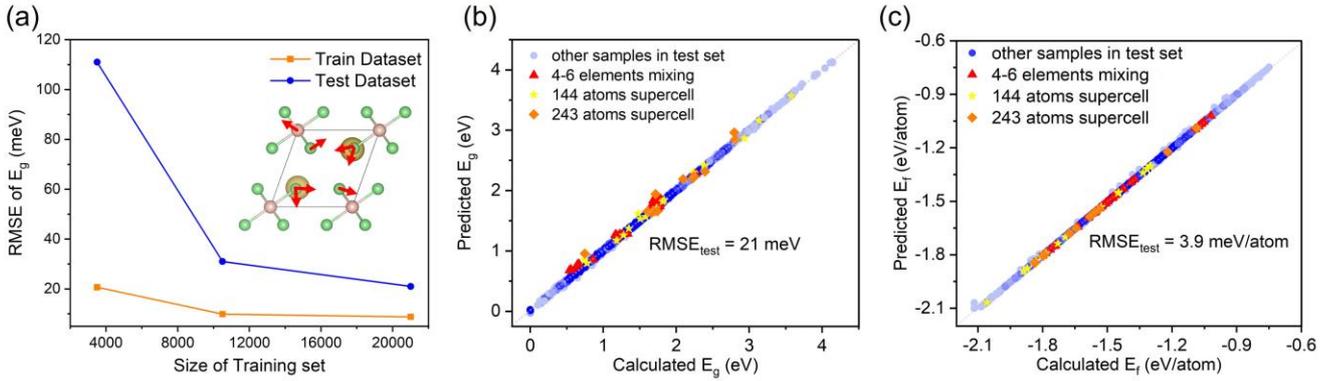

Fig. 6. (a) RMSE of the GNN model for predicting the DFT band gaps versus the size of augmented datasets. The inset shows the augmented data samples are created by applying random perturbations to the unrelaxed structures. The comparison between ML-predicted and DFT-calculated (b) band gaps and (c) formation energies of the samples in the test dataset.

TABLE 3. GNN model's RMSE and MAE of DFT band gap and formation energy for train dataset and test datasets. There are 3889 computed data and 23334 augmented data in the dataset.

|  | Train set | Test sets | | |
| --- | --- | --- | --- | --- |
| Details | 90% of the augmented dataset | 10% of the augmented dataset | 4 to 6 elements mixing, 72-atom supercells | 2 and 3 elements mixing, 144-atom supercell | 2 and 3 elements mixing, 243-atom supercell |
| Size of samples | 21000 | 2334 | 100 | 18 | 12 |
| RMSE of $E_g$ (meV) | 8.8 | 10.2 | 81 | 50 | 112 |
| MAE of $E_g$ (meV) | 6.3 | 7.2 | 71 | 34 | 87 |
| RMSE of $E_f$ (meV/atom) | 4.1 | 3.9 | 3.5 | 4.5 | 6.4 |
| MAE of $E_f$ (meV/atom) | 2.1 | 2.1 | 3.5 | 4.1 | 6.0 |

Table 3 presents a comprehensive evaluation of the performance of our model across various types of mixed VODPs. Utilizing a training dataset that encompasses systems with up to three mixing elements and supercells containing as many as 72 atoms, the GNN model demonstrates satisfactory performance on the test set, which includes structures featuring 4 to 6 elements mixed at the M-site or supercells comprising 144 and 243 atoms. The RMSE for the band gaps and formation energies in these more complex systems (with 4 to 6 mixing elements and larger supercells) is generally less than 0.1 eV and 7 meV/atom, respectively. This indicates that our model can be effectively applied to larger high-entropy mixed VODP supercells not included in the training dataset. Table 4 summarizes key findings from previous studies that employed machine learning methods for predicting band gaps in mixed perovskites. Compared to prior work, our approach not only exhibits competitive predictive performance but also highlights the applicability of our GNN model for modeling medium- and high-entropy systems characterized by four to six principal mixing elements.

TABLE 4. Performance of ML algorithms reported in literature toward band gaps of mixed perovskites.[30, 31, 32, 47, 48, 49]

| Systems | Details | ML model | Errors in $E_g$ (meV) |
| --- | --- | --- | --- |
| $A_2MX_6$ [this work] (A∈{Rb, Cs}, M∈{Ti, Zr, Pd, Sn, Te, Hf, Pt}, X∈{Cl, Br, I}) | A-site, M-site, X-site, AM-site, AX-site, MX-site or AMX-site mixing; up to 6 elements mixing | Transformer-inspired graph neural network | RMSE = 21 MAE = 10 |

| ABX$_3$ [32] (A∈{Cs, K, Rb}, B∈{Cd, Ge, Hg, Pb, Sn, Zn}, X∈{Cl, Br, I}) | B-site mixing; up to 4 elements mixing | Crystal graph convolution neural network | MAE = 37 |
|---|---|---|---|
| ABX$_3$ [30] (A∈{Cs, Rb, MA, FA}, B∈{Pb, Sn, Cd, Ge}, X∈{Cl, Br, I}) | A-site, B-site, X-site, AB-site, AX-site, BX-site or ABX-site mixing; up to 3 elements mixing | Extensive deep neural network | RMSE = 90 MAE = 69 |
| ABX$_3$ [31] (A∈{K, Rb, Cs, MA, FA}, B∈{Ca, Sr, Ba, Ge, Sn, Pb}, X∈{I, Br, Cl}) | B-site mixing; up to 4 elements mixing | Neural networks | RMSE = 220 |
| ABX$_3$ and Cs$_2$Ag{Sb, Bi}Br$_6$ [48] (A∈{Cs, Ba, Sr}, B∈{Pb, Sn, Ge}, X∈{I, Br, O}) | A-site, B-site or X-site mixing; binary mixing | SISSO | RMSE = 330 |
| ABX$_3$ [49] (A∈{Na, K, Rb, Cs, AM, HZ, MA, FA, DMA, EA, GUA}, B∈{Sn, Pb}, X∈{Br, I}) | A-site mixing; binary mixing | XGBoost models | MAE = 120 |
| ABX$_3$ [47] (A∈{Cs, Rb, K, Na}, B∈{Sn, Ge}, X∈{I, Br, Cl}) | A-site, B-site or X-site mixing; binary mixing | Kernel ridge regression model | RMSE = 146 |

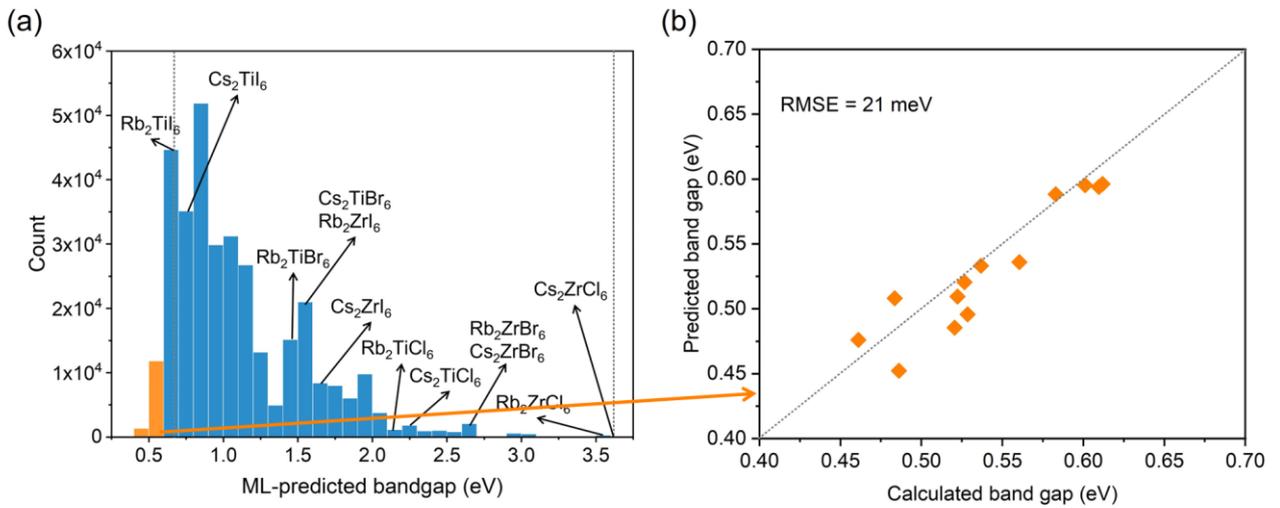

Fig. 7. (a) Distribution of the predicted band gaps of {CsRb}$_2${TiZr}{ClBrI}$_6$. The orange region represents the mixed VODPs with smaller band gaps than the pure systems. (b) Comparison between ML-predicted and DFT-calculated band gap for randomly selected mixed VODPs that have smaller band gaps than the pristine systems.

Next, we applied our GNN model to investigate selected mixed VODP systems. In typical mixed VODP systems, the band gap of a mixed material can be lower than the minimum band gap of the corresponding component materials. To assess whether the GNN model can capture this characteristic feature of mixed VODPs, we utilized the trained GNN model to predict the band gaps of all possible {CsRb}$_2${TiZr}{ClBrI}$_6$ systems simulated with 36-atoms supercells. The results indicate that certain mixed configurations exhibit notably smaller band gaps (illustrated as orange histograms in Fig. 7a) compared to the pure VODP counterparts. Subsequently, we select 14 samples that possess smaller band gaps than the pure VODPs to evaluate whether the ML-predicted band gaps align well with the DFT-calculated band gap. As depicted in Fig. 7b, the RMSE of the predicted bandgaps for these samples is approximately 25 meV, underscoring the accuracy of our model in capturing the properties of these outliers.

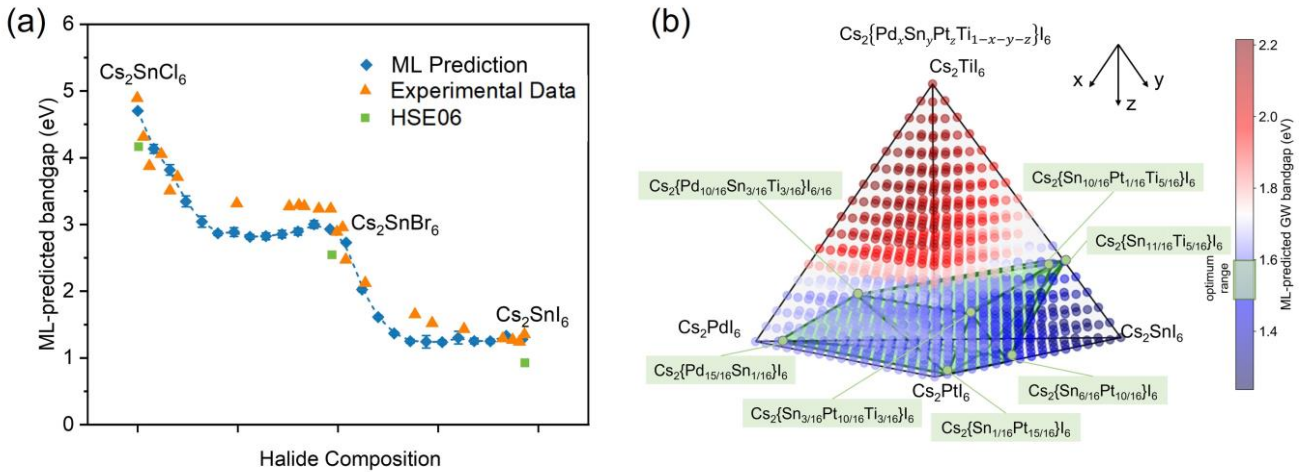

Fig. 8. (a) ML-predicted GW band gap for Cs$_2$Sn{ClBr}$_6$. Experimental data and calculations with the HSE06 functional[40] are shown in orange triangles and green squares for comparison. (b) The ML-predicted GW band gap for Cs$_2${Pd$_x$Sn$_y$Pt$_z$Ti$_{1-x-y-z}$}I$_6$. Inside the tetrahedron, the blue polygon shows the region of mixing ratios with which the mixed systems have band gap between 1.5 to 1.6 eV.

Previous experiments have shown that the band gap bowing effects can lead to mixed Cs$_2$SnX$_6$ having considerably lower values than those linearly extrapolated from their corresponding pure halide compounds[40]. To examine if our model can reproduce experimental finding, we compared experimental data (represented by orange rectangles) with the ML-predicted DFT band gap plus the GW corrections for $Cs_2Sn\{Cl_xBr_{1-x}\}_6$ and $Cs_2Sn\{I_xBr_{1-x}\}_6$ as shown in Fig. 8a. Herein, predicted DFT band gaps are estimated by averaging over 20 random mixing configurations comprising 144 atoms each; GW corrections are derived through the linear fitting using $E_g^{GW} = 1.46 E_g^{PBEsol} + 1.18$ (as obtained in Fig. 1c). Comparing the predicted quasiparticle band gaps against experimental optical band gaps effectively, we find our results well reproduce the experimental trend of optical band gaps and are even more accurate than the calculated results with the HSE06 functional[40]. The discrepancies between our predictions and the experiments partially comes from the exciton binding energies of Sn-based VODPs, which are in range of 130 ~ 730 meV)[50, 51, 52].

Our GNN model enables efficient screening of a substantial number of mixing VODPs to identify candidate materials with target band gaps, thereby providing useful references for experimental investigations. We conducted high-throughput predictions of the GW band gap of quaternary mixing VODPs $Cs_2\{Pd_xSn_yPt_zTi_{1-x-y-z}\}I_6$. The band gaps corresponding to each mixing ratio were estimated by averaging the results from 20 randomly generated supercell structures with the same fixed mixing ratio. Using a NVIDIA 3080 GPU, our high-throughput screening procedure required approximately 8 seconds to evaluate the band gaps of 19,380 mixed VODP supercells containing 144 atoms each—an endeavor that is challenging to achieve efficiently using conventional first-principles methods. As shown in Fig. 8b, each colored point denotes the predicted band gap of either a pure or mixed VODP with a given mixing ratio. In this figure, the blue polygon covers the mixing ratios of $Cs_2\{Pd_xSn_yPt_zTi_{1-x-y-z}\}I_6$ that fall within the optimal range of band gaps (1.5 ~ 1.6 eV) suitable for photovoltaic applications.

In summary, we investigated the diverse electronic and optical properties of VODPs and proposed a GNN model to capture nonlinear mixing effects across the vast chemical space of mixed VODPs. The GNN model leverages atomic-site occupation information to uniquely characterize structures and directly correlates them with their physical properties. It achieves a RMSE of 21 meV for band gaps and 3.9 meV/atom for formation energies on our test dataset. Furthermore, we demonstrate that data augmentation is an effective strategy for training our model to learn the "many-to-one" mapping from unrelaxed structures to their respective physical properties. Our study lays the groundwork for generalizing GNN-based models capable of accurately and efficiently describing medium- and high-entropy mixed perovskites comprising four to six principal mixing elements while utilizing a relatively small training set consisting of fewer than 4000 DFT calculations involving supercells with less than 80 atoms. Our GNN model also faithfully describes the experimentally observed nonlinear mixing effects in mixed VODPs, which can lead to notably smaller band gaps of mixed VODPs compared to the linear extrapolations of pristine systems. Finally, we apply our GNN model to screen mixed VODPs from an extensive dataset with target band gaps for future development of optoelectronic materials.

**Methods**

**First-principles calculations on ground-states and excited-states properties**

Density functional theory (DFT) calculations are carried out using plane-wave based methods implemented in QUANTUM ESPRESSO[53]. Optimized norm-conserving Vanderbilt pseudopotentials[54, 55] and PBEsol exchange correlation functional[56] are used in our calculations. The Kohn-Sham wavefunctions are expanded with a plane-wave basis set with a cutoff energy of 80 Ry. All structures are optimized until the total energy converges within $10^{-6}$ Ry and the forces are less than $10^{-4}$ Ry/Å. Spin-orbit coupling (SOC) is not considered in our calculations.

The quasiparticle energies are calculated using the one-shot GW approximation ($G_0W_0$ approximation), which is implemented in the BERKELEYGW package[57]. We use the Hybertsen-Louie generalized plasmon pole model to treat the frequency-dependent dielectric function[58]. Our calculations include 1000 empty states in combination with the static-remainder approach for the quasiparticle properties. The dielectric matrices are calculated with plane-wave basis with a 35 Ry kinetic-energy cutoff.

To calculated the excitonic properties, the Bethe-Salpeter equation (BSE) are solved within the Tamm-Dancoff approximation[59, 60], as implemented in the BERKELEYGW package[57]:

$$\left(E_{ck}^{QP} - E_{vk}^{QP}\right)A_{vck}^S + \sum_{v'c'k'} \langle vck|K^{eh}|v'c'k'\rangle = E_{ex}^S A_{vck}^S$$

where $E_{nk}^{QP}$ are the GW quasiparticle energies, $A_{vck}^S$ is the eigenvector of the excitonic state $S$, $E_{ex}^S$ is the exciton energy, and $K^{eh}$ is the electron–hole (e-h) interaction kernel matrix. The e-h interaction kernel matrix is first calculated using a 4 × 4 × 4 k-grid, considering 9 valence and 5 conduction bands for $Cs_2TiI_6$, 15 valence and 5 conduction bands for $Cs_2TeI_6$ and 14 valence and 7 conduction bands for Cs$_2$SnI$_6$. The results are then interpolated onto a finer 8 × 8 × 8 k-grid.

**Details of the GNN model**

In the following discussions, the meanings of common variable notations are listed in Table 5. Consider an unrelaxed supercell with given chemical compositions and atomic occupations, the coordinates of the atoms within the supercell are $r_i$ ($i = 1, \ldots, N_{atoms}$). With the unrelaxed structure, one can build a multi-edge graph that encodes the relative positions of atoms in the infinite crystal. Node $i$ in a multi-edge graph corresponds to an atom and all its periodic images in the crystal. Each node $i$ has an invariant embedding vector $\boldsymbol{a_i} \in \mathbb{R}^{k_a}$ ($k_a = 19$) and an equivariant embedding vector $\boldsymbol{\chi_i} \in \mathbb{R}^3$. The invariant embedding vector $\boldsymbol{a_i}$ is initialized by a linear transformation of the atomic information containing the electron shell configuration, electronegativity, ionic radius and the element of site $i$ (Fig. 9): $\boldsymbol{a_i} \leftarrow \mathbf{W} \cdot \mathbf{AtomicInformation}(i) + b$, where the weight matrix $\mathbf{W}$ and the bias $b$ are trainable parameters. The equivariant embedding vector $\boldsymbol{\chi_i}$ is initialized with zeros and updated in the forward pass of the graph neural network model.

TABLE 5. Meaning of variable notations in our discussion

| Notations | Meaning |
| --- | --- |
| Bold, italic, lower-case symbols | Vectors |
| Bold, upper-case symbols | Matrices |
| Regular, italic symbols | Scalars, functions |
| $|\boldsymbol{r}|$ | 2-norm of a vector $\boldsymbol{r}$ |
| $Ln^K$, $Ln^Q$, $Ln^V$, $Ln$ | Linear transformation |
| $\phi_\alpha$, $\phi_\beta$, $\phi_{msg}$, $\sigma$, $\phi_{out}$ | Multi-layer perceptron |
| $\circ$ | Element-wise multiplication |
| $|$ | Concatenation of vectors |
| $\mathrm{MeanPool}_{i \in G}$ ($\mathrm{MaxPool}_{i \in G}$) | Perform mean (max) pooling operation over all nodes $i$ in graph $G$ |
| Flatten | Reshape a tensor to a row vector |
| Sigmoid(x) | Sigmoid function $\frac{1}{1+e^{-x}}$ |

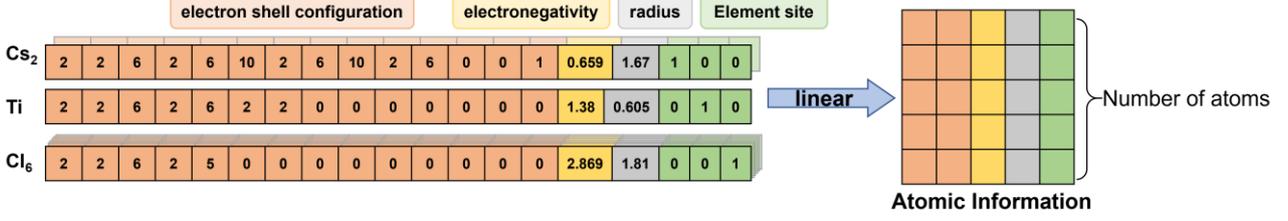

Fig. 9. The invariant node embedding vector $a_i$.

An edge indexed with $(i, j, g)$ is constructed from node $j$ to node $i$, if a periodic image located at $r_{j,g} = r_j + R_g$ represented by node $j$ is one of the $K_i$ nearest neighbors of atom at $r_{i,0} = r_i$, where the numbers of neighbors $K_i$ considered for different sites $i$ are listed in Table 2 and $R_g$ is a lattice vector. The distance vector corresponds to edge $(i, j, g)$ is $r_{ij,g} = r_{j,g} - r_i$. Each edge has an equivariant edge embedding vector initialized with $\bar{r}_{ij,g} = r_{ij,g}/\langle r \rangle$ and an invariant embedding vector $e_{ij,g}$, which is initialized using a radial basis function (RBF) expansion of $\bar{r}_{ij,g}$. Here $\langle r \rangle = \frac{1}{N_{edge}} \sum_{ij,g} |r_{ij,g}|$ is the average length of the distance vectors within a graph. The RBFs are gaussian functions. Unlike other graph neural networks which use exact structures as input, the edge embedding vectors $\bar{r}_{ij,g}$ and $e_{ij,g}$ are less dependent on the exact bond length, since they are normalized with $\langle r \rangle$. In addition, the embedding vectors and our model layers are invariant with the choices of supercells, because the lattice vector basis are not explicitly used in our model.

**Transformer-like Graph Layer**

The Transformer-like Graph Layer (TGL) updates the node embeddings with the following steps:

(1) Calculate intermediate variables key, query, and values:

$$k_{ij,g} = Ln^K(a_i)|Ln^E(e_{ij,g}), q_i = Ln^Q(a_i), v_{ij,g} = Ln^V(a_i)|Ln^V(a_j)|Ln^E(e_{ij,g})$$

(2) Calculate messages $msg_{ij,g}$ via edge $(i, j, g)$ and aggregate messages from the neighbors $j \in N(i)$, where $N(i)$ represents all the neighboring nodes of node $i$

$$w_{ij,g} = \text{Sigmoid}(\frac{1}{d_q} q_i \circ \sigma(k_{ij,g})),$$

$$\Rightarrow msg_{ij,g} = w_{ij} \circ \phi_{\text{msg}}(v_{ij}),$$

$$\Rightarrow msg_i = \frac{1}{D_i} \sum_{j \in N(i)} \sum_g w_{ij,g} \circ msg_{ij,g},$$

where $D_i$ is the number of edges that have the destination node $i$ and $d_q$ is the dimension of vector $q_i$. Then we calculate the coefficient $\alpha_{ij,g}$ and $\beta_{ij,g}$, which are used to update the equivariant node embedding vectors

$$\alpha_{ij,g} = \phi_\alpha([\bar{r}_{ij,g} \cdot \chi_i, \bar{r}_{ij,g} \cdot \chi_i, \chi_i \cdot \chi_j, |\bar{r}_{ij,g}|, |\chi_i|, |\chi_j|]|a_i|a_j),$$

$$\beta_{ij,g} = \text{Sigmoid}\left(\phi_\beta([\bar{r}_{ij,g} \cdot \chi_i, \bar{r}_{ij,g} \cdot \chi_i, \chi_i \cdot \chi_j, |\bar{r}_{ij,g}|, |\chi_i|, |\chi_j|]|a_i|a_j)\right).$$

Note that both $\alpha_{ij,g}$ and $\beta_{ij,g}$ are invariant with the rotation of coordinates because the calculations take dot products between vectors as inputs.

(3) Update the invariant and equivariant node embedding vector of node $i$

$$a_i \leftarrow \phi_\text{out}(a_i|msg_i)$$

$$\chi_i \leftarrow \frac{1}{D_i}\sum_{j\in N(i)}(\alpha_{ij}r_{ij} + \beta_{ij}\chi_j).$$

After each TGL layer, we perform graph-level pooling operations to collect the information of $a_i$ and $\chi_i$. Here we use a superscript "$(m)$" to mark the variables in the $m$-th TGL layer:

$$h^{(m)} = [\text{MaxPool}_{i\in G}\left(\psi_1\left(a_i^{(m)}\right)\right), \text{MeanPool}_{i\in G}\left(\psi_2\left(a_i^{(m)}\right)\right)],$$

$$\lambda^{(m)} = \text{MeanPool}_{i\in G}\left(\chi_i^{(m)}\right).$$

To predict the final result, such as the band gap and formation energy, we use a multi-layer perceptron to process the information encoded in the graph-level descriptors $\lambda^{(m)}$ and $h^{(m)}$:

$$\Lambda = [\lambda^{(1)}, \lambda^{(2)}, \ldots]; \mathbf{H} = [h^{(1)}, h^{(2)}, \ldots],$$

$$\hat{y} = \text{MLP}\left(\text{Flatten}(\Lambda^\mathbf{T}\Lambda)\middle|\text{Flatten}(\mathbf{H})\right).$$

**Model Training**

For model training, we use the ADAM optimizer with a learning rate of 0.0015 and a batch size ranging from 50 to 400. The batch size does not affect the training results significantly. Our model was implemented with the Pytorch and PyG (PyTorch Geometric) ML frameworks. In addition, we also optimize the hyperparameters of our GNN model, which is shown in Fig. S5. For the results given in the paper, the model uses 3 TGL layers and 64 nodes of hidden layers with the minimum RMSE after 40,000 epochs. The number of parameters of this model is around $9 \times 10^4$.

## Data availability

All data used in this work is available at Figshare link https://figshare.com/articles/dataset/mixed_VODPs_dataset/27752469.

## Code availability

The code and full model used in this work are available on GitHub at https://github.com/ZhangFan-phy/VODP-GNN.

## Acknowledgements

We acknowledge the support by the National Natural Science Foundation of China (12474221 and 12104080), the Guangdong Provincial Quantum Science Strategic Initiative (Grant No. GDZX2401002), the Fundamental Research Funds for the Central Universities (DUT24LK007), and GHfund A (202407011848). Computational resources are provided by the National Supercomputer Center at Wuzhen.

## Author contributions

F. Z. and W. G. designed the workflow of the project and developed the Python code for GNN model. F. Z. performed the DFT calculations and trained the ML models. L. F. helped in the interpretation of the work. P. Z. and J. Z. supervised the work. All authors modified and discussed the paper together.